\documentclass[]{interact}

\usepackage[natbibapa,nodoi]{apacite}
\usepackage{epstopdf}
\usepackage[caption=false]{subfig}

\usepackage[longnamesfirst,sort]{natbib}
\bibpunct[, ]{(}{)}{;}{a}{,}{,}
\renewcommand\bibfont{\fontsize{10}{12}\selectfont}

\theoremstyle{plain}

\theoremstyle{definition}

\theoremstyle{remark}

\begin{document}

\articletype{ARTICLE TEMPLATE}

\title{Projected gradient descent–based optimization of 3D bobsleigh track centerlines from 2D data for simulation}

\author{
\name{Zhe Chen\textsuperscript{a}\thanks{CONTACT Jicheng Chen Email: jichengc@buaa.edu.cn}, Huichao Zhao\textsuperscript{b}, Yongfeng Jiang\textsuperscript{b}, Minghui Bai\textsuperscript{b}, Lun Li\textsuperscript{b} and Jicheng Chen\textsuperscript{c}}
\affil{\textsuperscript{a}School of Tansportation Science and Engineering, Beihang University, Beijing,  China; \textsuperscript{b}FAW R\&D General Institute, Changchun, China; \textsuperscript{c}School of Reliability and Systems Engineering, Beihang University, Beijing,  China}
}

\maketitle

\begin{abstract}
This paper proposes a method for generating a 3-dimensional (3D) track centerline based on 2D centerline data. Incorporating international track design regulations, the method formulates an optimization problem that considers total track length, height difference, slope constraints, and geometric continuity. A Projected Gradient Descent (PGD) algorithm is used to solve the optimization problem. Within the selected trajectories and parameter ranges in this study, the proposed algorithm can compute centerlines whose characteristic parameter trends are consistent with those of the actual track, based on either actual or scaled 2D data. Compared with the actual track, the maximum errors ranges for length, height difference, and average gradient are 1.1\%, 16.4\%, and 13.5\%. Track sensitivity to segmentation and height-difference weighting varies with data type, and appropriate selection of segmentation and weighting parameters effectively reduces deviations and enhances the geometric accuracy of the generated centerlines. 
\end{abstract}


\section{Introduction}

Since bobsleigh was formally recognized as a Winter Olympics event in 1924, it has attracted widespread attention from participating countries \citep{bi:bu4}. Bobsleighs are unpowered vehicles that rely entirely on the initial push from athletes and the conversion of gravitational potential energy into kinetic energy as they descend the track \citep{bi:1, 3, 5}. The competition is timed to an accuracy of milliseconds, and the final times often differ by only a few tenths of a second\citep{4}. To minimize energy loss, athletes typically avoid using the brake unless necessary before crossing the finish line. The dynamics of the bobsleigh and the driver's control skill are key factors influencing performance \citep{bi:bu1, roberts2013}.

However, due to restrictions from official governing bodies and the high cost of maintaining bobsleigh tracks\citep{7}, even countries with access to tracks often do not allow athletes to train on them regularly. Typically, limited test runs are permitted only shortly before official competitions \citep{bi:2, rempfler2015, 4, 6}. Consequently, simulation-based training and performance analysis have become vital tools for optimizing bobsleigh design and improving athlete technique. Researchers have developed virtual tracks, bobsleigh-blade/ice interaction models, multibody dynamics models of bobsleighs, and driver models. These simulation frameworks are increasingly being used to support bobsleigh structural design and help athletes devise optimal racing strategies \citep{bi:2}. Besides, there are limited studies on optimizing the centerlines of bobsleigh tracks.

Early bobsleigh simulation research was significantly constrained by the limited computational capabilities of CPUs at the time, often relying on simplified point-mass models moving on smooth hyperbolic surfaces \citep{bi:3, bi:4}. The study \citep{bi:4} developed a partial surface model of a bobsleigh track, treating the bobsleigh as a point mass to analyze its dynamic response under external forces. However, the authors also acknowledged that due to the lack of available data, virtual training environments differ significantly from the real settings, limiting the variety of technical strategies athletes can efficiently practice. Based on this point-mass framework and real data from the Lillehammer track in Norway, The study \citep{bi:5} further designed an optimal controller to simulate bobsleigh driving behavior, aiding in athlete training. To better capture the complex nonlinear dynamics of bobsleigh motion on narrow, twisted, banked, and icy tracks, a geometric model is proposed \citep{bi:6}, which represents the tracks as a sequence of discrete points arranged along a 3D spatial curve. Moreover, a comprehensive simulation framework incorporating a blade–ice contact friction model is designed and validated on the Cesana and Whistler tracks, offering valuable guidance for track design and athlete training. The study \citep{bi:8} further validates its agreement between the simulated and experimental results on the Cesana track. However, the study did not involve real-time simulation, and the applicability of the proposed model to other official tracks worldwide was not examined. Subsequently, the geometric model is further extended into a multibody dynamics model \citep{bi:2}, where more detailed dynamic characteristics are considered, resulting in better driving fidelity. Recent advances in bobsleigh simulation have focused on improving blade–ice friction modeling and virtual training environments \citep{bi:bu3, bi:7}. The study \citep{bi:10} developed longitudinal and lateral friction models through experiments, FEM and nonlinear regression, validating their effectiveness in simulators to evaluate driving styles. Some studies further integrate steering and visual feedback or design immersive systems for training and tourism \citep{bi:2, bi:3,bi:10, bi:11}. The study \citep{bi:11} introduced a VR hammock system to mimic sliding sensations, while the studies \citep{bi:12, bi:13} proposed automated methods for generating 3D track surfaces.

However, most of the aforementioned studies focus on the simulation of only one or two tracks, and the conclusions drawn lack generalizability. Many researchers have acknowledged the necessity of validating their proposed models against simulation results from other tracks in order to derive more universally applicable insights for athlete training. Currently, track data are primarily obtained through a combination of field measurements and industrial software modeling, which involves significant labor and material costs. Moreover, most countries do not publicly disclose the official data of their bobsleigh tracks. Therefore, acquiring a greater variety of available official track data worldwide has become an urgent need for enhancing both the diversity of research findings in simulation analysis and the general applicability of virtual training outcomes for bobsleigh athletes. The centerline of a bobsleigh track serves as a crucial dataset that defines a trajectory of a track. In contrast, the cross-sectional shapes are relatively standardized and can be generated using industrial software based on common templates \citep{bi:12}. 

This study focuses on typical bobsleigh track centerlines that determine the trajectory characteristics of the track and proposes a method for generating 3D centerlines from 2D centerlines. In accordance with international track design regulations, an optimization problem is formulated that considers factors such as total track length, height difference, slope, and continuity. And a PGD algorithm is used to solve the problem. The generated 3D centerline is quantitatively compared with actual track data, and the influence of different track characteristics and algorithmic parameters on the generation results is investigated. The proposed method provides data support for enhancing the diversity of theoretical research and improving the generalizability of training outcomes for bobsleigh athletes.

\section{Methodology and Computational Framework}

\subsection{Formulation of the optimization problem}
According to the International Olympic Bobsleigh Competition Rules \citep{bi:14}, a bobsleigh track consists of a sequence of straight and curved sections. These two types of track segments exhibit distinct geometric characteristics, which necessitate different approaches for their analysis and generation. Therefore, it is essential to segment the track into straight and curved sections. Since the curvature of a straight segment is zero, curvature serves as a key variable for distinguishing between straight and curved portions. 

\begin{figure}
\centering
\resizebox*{13cm}{!}{\includegraphics{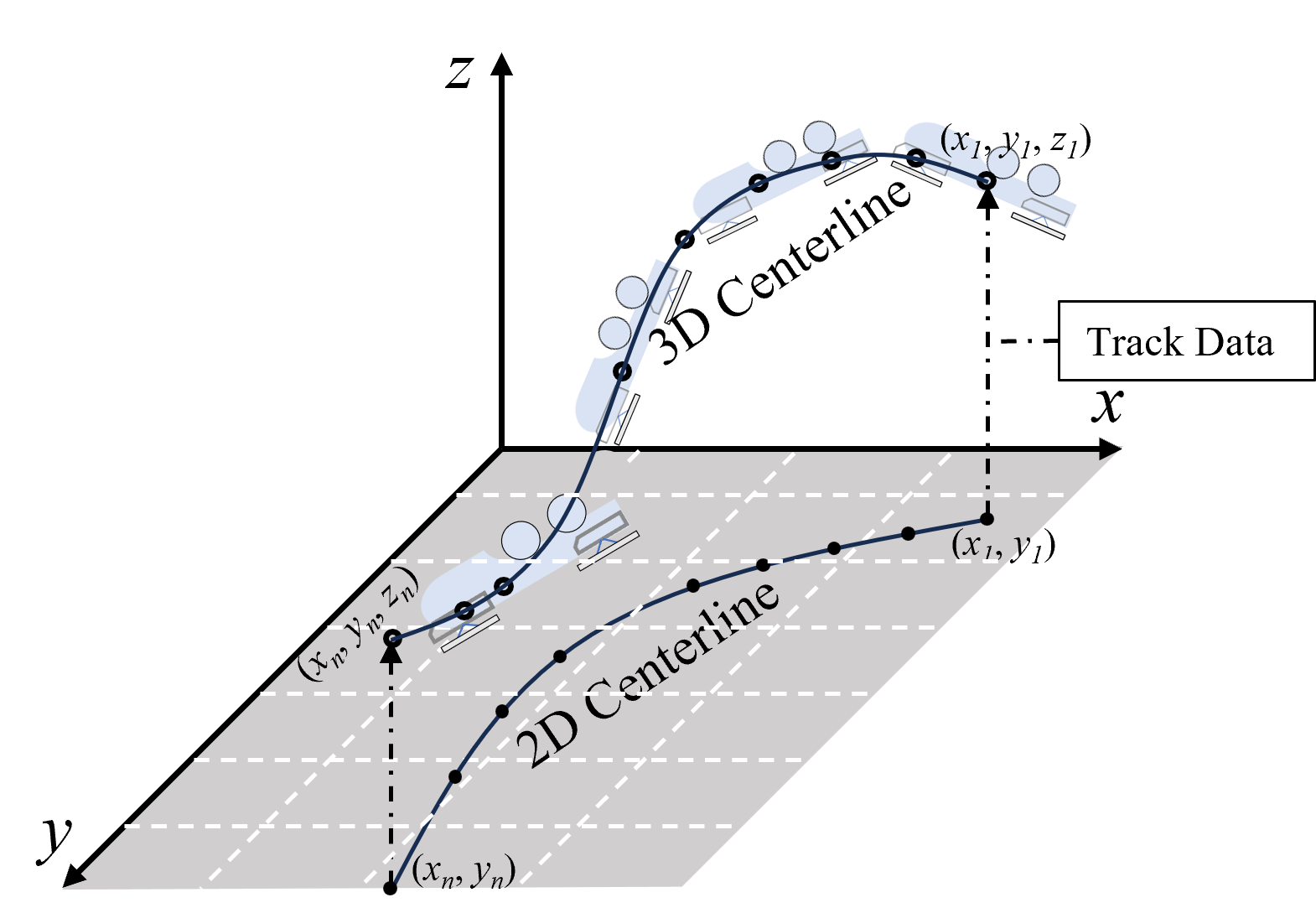}}\hspace{5pt}
\caption{ Transition from a 2D centerline to a 3D centerline.} \label{figure1}
\end{figure}

The horizontal plane is defined by the $x$ and $y$ axes, while the $z$-axis is oriented perpendicular to this plane. Taking the starting point of the track as the origin, a set of equally spaced points are sampled along the 2D centerline. The generation is shown in Figure~\ref{figure1}. The 2D curvature at each point is then calculated by
\begin{equation}
		K_{pk} = \frac{|\ddot{x} \dot{y} - \dot{x} \ddot{y}|}{\left( \dot{x}^2 + \dot{y}^2 \right)^{3/2}} 
	\end{equation}
Where $K_{pk}$ denotes the curvature at the current point, and $x$ and $y$ represent the coordinates of the point on the $x$-and $y$-axes, respectively. $\dot{x}$ and $\dot{y}$ are the first-order finite differences of the $x$ and $y$ coordinates at the current point, while $\ddot{x}$ and $\ddot{y}$ represent the corresponding second-order finite differences.  

According to the calculated curvature values, points with curvature below a predefined threshold are classified as points on straight segments. However, due to potential measurement errors in the raw data, individual points may be misclassified. To address this issue, the classification of each point is determined based on the classification of its neighboring region. Specifically, if the neighboring points are identified as part of a straight segment and the curvature at the current point is also below the threshold, the point is considered to lie on a straight segment. Otherwise, it is classified as part of a curved segment. 

For bobsleigh tracks, slope and length are two critical characteristics. For a straight segment, the slope $g_k$ remains constant throughout. For a curved segment, when the distance $d$ between two points is extremely small, the segment between them can be approximated as a straight line.

Therefore, if the coordinates of the previous point are $(x_{k-1}, y_{k-1}, z_{k-1})$, and the current point’s $x$ and $y$ coordinates are $(x_k, y_k)$, then the $z$-coordinate of the current point can be approximated as $z_k= z_{k-1} - d \cdot g_k$. The distance $d$ is calculated as follows:
	\begin{equation}
		d = \sqrt{(x_k - x_{k-1})^2 + (y_k - y_{k-1})^2}
	\end{equation}
	
The starting point of the track is defined as $(0, 0, H)$, where $H$ represents the height difference along the $z$-axis between the highest point at the start and the lowest point of the entire track. Therefore, it is necessary to determine the height and slope of each track segment, which allows the 3D coordinates of each point to be recursively computed.

The generated 3D coordinates should satisfy several constraints:
\begin{enumerate}
\item The slope of each segment must remain within a reasonable range;

\item The total length of the generated 3D track should be equal to the known total track length $L$;

\item The total height difference of the generated track must match the known value $H$;

\item To ensure the continuity of the bobsleigh track, the curvature in the $z$-axis direction should be significantly smaller than that in the $x$ and $y$ axes, meaning that the 2D curvature at each point should approximately equal the 3D curvature;

\item The average slope of the track should be equal to the known average slope $\overline{g_k}$.

\end{enumerate}

In summary, suppose the 2D coordinates of the track consist of $n$ points $(x_k, y_k)$, where $k = 1, 2, \ldots, n$, and the 2D curvature at each point is denoted by $K_{pk}$. Let the track be divided into $j$ segments, with the slope of each segment denoted by $g_m$, where $m = 1, 2, \ldots, j$. The generated 3D coordinates for each point are $(x_k, y_k, z_k)$, and the 3D curvature at each point is $K_{sk}$. Let $L$ be the actual total track length, $H$ the actual total height difference, and $\bar{g}$ the actual average slope.

The generation of the 3D track from the 2D centerline can then be formulated as an optimization problem, as shown in Equation~(\ref{eq:3}), in which the slope of each segment is optimized such that the resulting 3D coordinates satisfy the five aforementioned constraints.
\begin{equation}\label{eq:3}
		\begin{aligned}
			J = \min_{g_k} \Bigg[ 
			& a \left( L - \sum_{k=2}^{n} d_k \right)^2
			+ b \left( H - \sum_{k=2}^{n} \sqrt{(z_k - z_{k-1})^2} \right)^2 \\
			& + c \left(\frac{1} {n} \sum_{k=1}^{n} (K_{pk} - K_{sk}) \right)^2
			+ d \left( \frac{1}{j} \sum_{m=1}^{j} g_m - \bar{g} \right)^2
			\Bigg] \\
			& \text{s.t.} \quad g_{k_\text{min}} \leq g_k \leq g_{k_\text{max}}
		\end{aligned}
	\end{equation}
In this equation, $J$ represents the cost function to be minimized, and $j$ is the number of track segments. The parameters $a$, $b$, $c$, and $d$ are the weighting factors corresponding to the total length difference, height difference, curvature difference between the 2D and 3D centerlines, and the average slope difference, respectively, $g_{k_\text{max}}$ and $g_{k_\text{min}}$ denote the maximum and minimum allowable slopes, $d_k$ is the 3D distance between point $k$ and point $k-1$, calculated as follows:
	\begin{equation}
		d_k = \sqrt{(x_k - x_{k-1})^2 + (y_k - y_{k-1})^2 + (z_k - z_{k-1})^2}
	\end{equation}
	
The calculation formula for the 3D curvature $K_{sk}$ at each point is as follows:
\begin{equation}
		K_{sk} = \frac{\sqrt{ \lvert u \rvert^2 \lvert v \rvert^2 - (v u)^2 }}{\lvert v \rvert^3}
	\end{equation}
where $u = [\dot{x}, \dot{y}, \dot{z}]$ is the first derivative vector of the coordinates, and $v = [\ddot{x}, \ddot{y}, \ddot{z}]$ is the second derivative vector of the coordinates.
To solve the bobsleigh track optimization problem established, a PGD algorithm \citep{9, 10} is used for parameter optimization. Compared to more complex methods such as genetic algorithms or simulated annealing, gradient descent offers faster convergence and requires fewer hyperparameters, making it more practical for repeated evaluations during track generation. 

In this study, the height difference $h_m$ of each track segment is a high-dimensional parameter (with dimensionality corresponding to the number of track segments), which introduces significant nonlinearity to the optimization problem, making the exact computation. Therefore, the variation of the cost function is employed as a substitute for gradient computation. This approach does not rely on the explicit calculation of partial derivatives but instead approximates the gradient through a numerical difference method. The corresponding computation is given as follows:
\begin{equation}
		\frac{\partial J}{\partial h_i} \approx \frac{J(h_i) - J(h_{i-1})}{h_i - h_{i-1}}
	\end{equation}
	where $i$ denotes the iteration index.

If the optimization is based on 2D coordinate data with uniform scaling, an additional optimization variable, the scaling factor $f$, must be introduced. In this case, the corresponding computation is modified as follows:
	\begin{equation}
	\frac{\partial J}{\partial f_i} \approx \frac{J(f_i) - J(f_{i-1})}{f_i - f_{i-1}}
	\end{equation}
	
	The gradient descent update rule is given as follows:
	\begin{equation}
	h_{i+1} = h_i - \eta \frac{\partial J}{\partial h_i}
	\end{equation}
	where $\eta$ is the learning rate, which is set to $10^{-5}$.
	If the scaling factor is considered, it should also be updated using gradient descent as follows:
	\begin{equation}
	f_{i+1} = f_i - \eta \frac{\partial J}{\partial f_i}
	\end{equation}
	
Compared to SGD (Stochastic Gradient Descent) algorithms \citep{11, 12, 13}, PGD algorithms can be regarded as projecting the height difference of each segment during optimization to ensure that the corresponding slope of each segment falls within the prescribed limits, which in turn satisfies the overall constraint conditions. Although the projection is not explicitly performed on each variable (e.g., the height difference of each segment) at every gradient descent step, it is effectively realized by adjusting the slope of each segment to meet the constraints.

During the gradient descent process, the height difference of the current segment is denoted as $h_m$, and the corresponding slope is calculated as:
	\begin{equation}
	g_m = \frac{h_m}{d_m}
	\end{equation}
	where $d_m$ represents the total length of a specific segment of the track, and it is calculated as follows:
	\begin{equation}
	d_m = \sum_{k = p_1 + p_2 + \cdots + p_{m-1}}^{p_1 + p_2 + \cdots + p_m}d_k
	\end{equation}
where $p_m$ denotes the number of points within segment $m$ of the track. The slope $g_m$ is constrained to remain within a predefined range. If $g_m$ exceeds this range (e.g., smaller than min slope or larger than max slope), the corresponding height difference $h_m$ is adjusted accordingly:
	\begin{equation}
	\left\{
	\begin{aligned}
	h_m &= g_{k_\text{min}} \times d_m, \quad && g_m \leq g_{k_\text{min}} \\
	h_m &= g_{k_\text{max}} \times d_m, \quad && g_m \geq g_{k_\text{max}}
	\end{aligned}
	\right.
	\end{equation}
	\begin{equation}
	\Delta J = |J_{i+1} - J_i| < \delta
	\end{equation}
	where $\delta$ represents the threshold of variation, which is set to $10^{-3}$.
\subsection{Numerical Procedure}

\begin{figure}
\centering
\resizebox*{8cm}{!}{\includegraphics{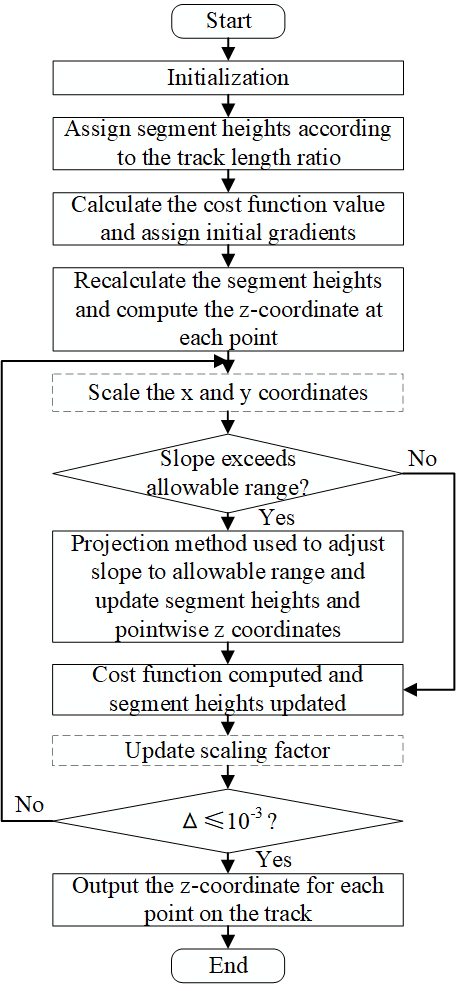}}\hspace{5pt}
\caption{ Numerical procedure.} \label{figure2}
\end{figure}

Based on the optimization equation formulated in accordance with the track design rules and the PGD algorithm, the computational procedure is illustrated in figure~\ref{figure2}. Firstly, the initial segmental height differences and the corresponding cost function value are calculated. Then, using a assigned initial gradient to update segmental height differences, and the $z$-coordinates of all points are recalculated based on these updated values.

Subsequently, the program enters the PGD algorithm iterative loop. At each iteration, it first checks whether the slope of each track segment falls within the specified bounds. If any segment's slope exceeds the upper or lower limits, the projection method is applied to map it back into the feasible range. Using the projected slope values, to update the segmental height differences, followed by recalculating the $z$-coordinates of all points.

The cost function is then evaluated again, and the height differences are further updated. This process repeats until the change in the cost function value converges to a predefined threshold. At that point, the final $z$ for all points on the track are determined.

As shown in the dashed box in the figure, if a scaling factor is included in the optimization, the $x$ and $y$ are scaled at the beginning of each iteration, and the scaling factor is updated at the end of each iteration.

\section{Model Verification}\label{sec:3.1}
	
	\subsection{Track with Real 2D Plane Data}\label{311}
Since some studies include real 2D plane data for a limited number of tracks \citep{bi:6}, this data can be used for solving the optimization problem. The generated track data are then compared with the actual track data from the studies. Due to the fact that the actual track height difference parameter that can be obtained can only be reflected in the descending section of the track, and the descending section of the track is the key factor affecting the competition results. Therefore, in this section, the impact of the track's final ascent sections on the results is not considered.

Table~\ref{table1} present comparisons of the geometric parameters of the generated and actual bobsleigh track centerlines for the Whistler and Cesana tracks. As shown in table~\ref{table1}, for the Whistler track, the generated tracks demonstrate strong consistency with the actual tracks in terms of total length, height difference, and average slope, with maximum deviations of 0.7\%, 0.4\%, and 0.5\%, respectively. For the Cesana track, the corresponding deviations are 0.2\%, 0.6\%, and 1.1\%. Furthermore, the generated tracks exhibit smaller average differences between the 2D and 3D curvature compared with their actual counterparts, while the slope values remain within the defined design limits.

		\begin{table}
		\caption{Comparison of geometric parameters of the generated and actual Whistler track.}
		\resizebox{\textwidth}{!}
		{\begin{tabular}{cccccc} 
				\toprule 
				& Total Length (m) & Height Difference (m) & Average difference between 2D and 3D curvature & Max Slope & Average Slope \\ 
				\midrule 
			Generated Track (Whistler) & 1286.7 & 148.7 & 3.97\texttimes10\textsuperscript{-4} & 0.152 &  0.1182\\
			Actual Track (Whistler) & 1278 & 149.3 & 9.76\texttimes10\textsuperscript{-4} &0.204  & 0.1176  \\ 
			Generated Track (Cesana) & 1356.7 & 123.7 & 2.52\texttimes10\textsuperscript{-4} & 0.147 &  0.0920\\
			Actual Track (Cesana) & 1359.2 & 124.5 & 5.04\texttimes10\textsuperscript{-4} &0.183  & 0.0910\\ 
			\bottomrule
		\end{tabular}}
		\label{table1}
	\end{table}
	
\begin{figure}
		\centering
		\subfloat[Whistler track.]{%
			\resizebox*{10cm}{!}{\includegraphics{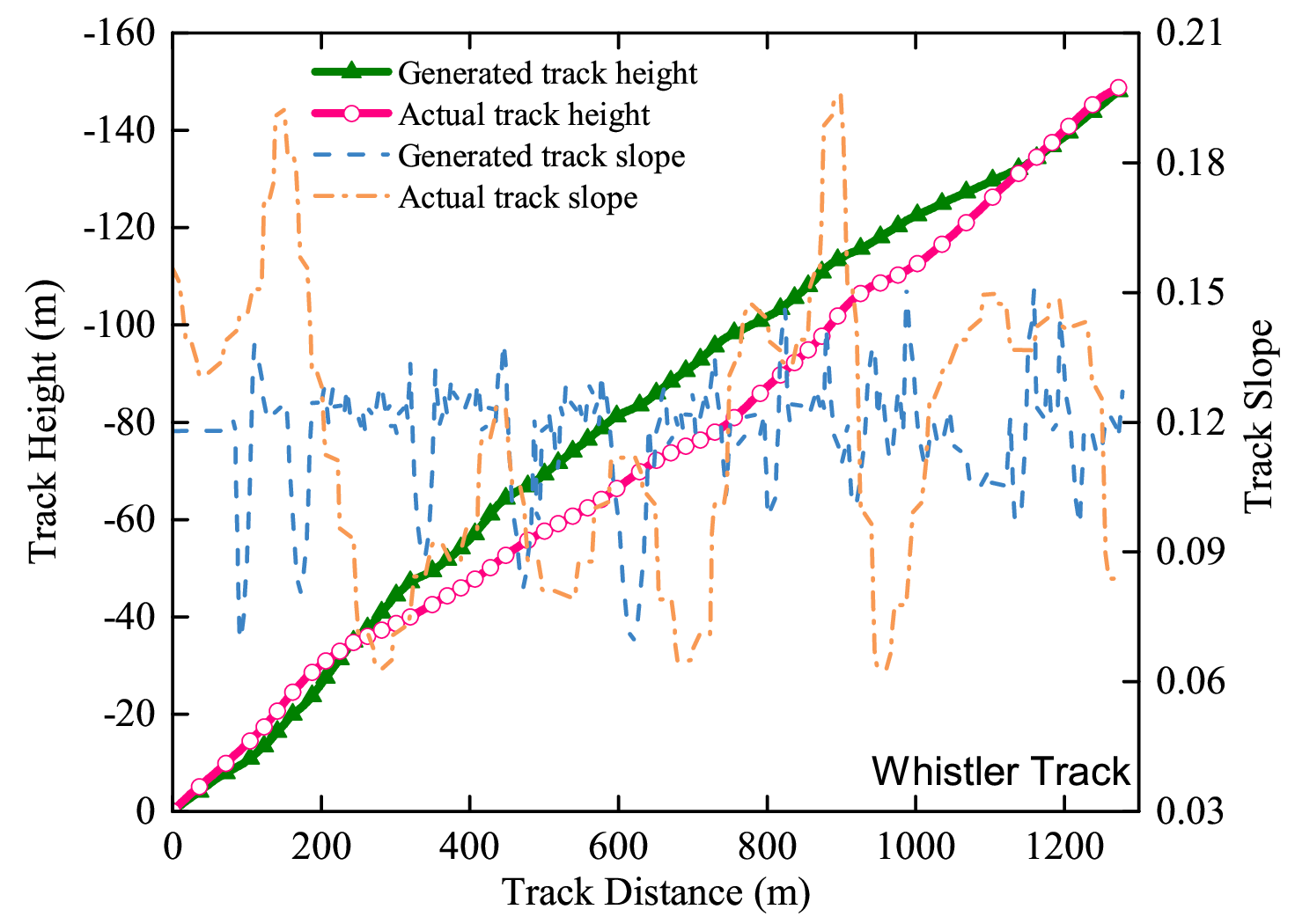}}}\hspace{5pt}
		\subfloat[Cesana track.]{%
			\resizebox*{10cm}{!}{\includegraphics{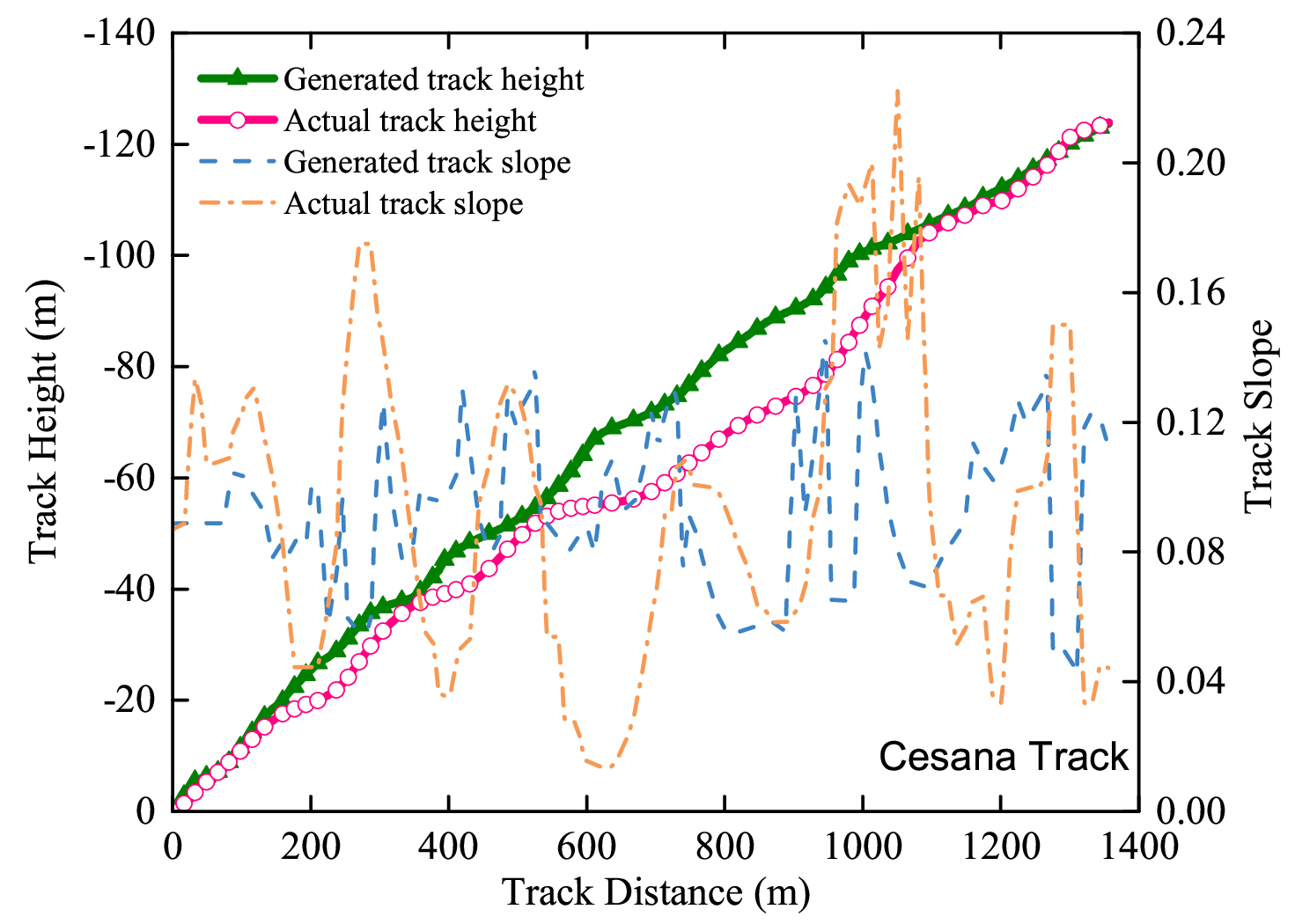}}}
		\caption{The comparison of the track height variation and slope along the distance for the generated and actual track centerlines.} \label{figure3}
	\end{figure}

	Figure~\ref{figure3} shows the comparison of the track height variation and slope along the distance for the generated and actual track centerlines. Among them, Figure~\ref{figure3}(a) is the Whistler track, and Figure~\ref{figure3}(b) is the Cesana track. 
	
	For the generated Whistler track, the height is lower than that of the actual track from 0 to approximately 245m, but higher between 245m and 1140m. Beyond 1140m, the heights converge. This is mainly attributed to the influence of slope variations. From the overall slope variation curve, the slopes of the generated track remain within the reasonable design range. The slopes of the actual and generated tracks alternate, resulting in height differences. For example, between 0 and 205m, the slope of the generated track is smaller than that of the actual track, leading to a lower height in this section. In contrast, from 205m to 245m, the slope of the generated track exceeds that of the actual track, causing the height to gradually approach that of the actual track.
	
	For the generated Cesana track, the height is close to that of the actual track from 0 to approximately 115m, but higher between 115m and 1266m. Beyond 1266m, the heights converge as well. In addition, due to slope variations, the slope of the generated track is greater than that of the actual track in most regions between 115m and 950m, resulting in higher track heights. In contrast, between 950m and 1110m, the slope of the generated track is smaller than that of the actual track, leading to a gradual convergence of height toward that of the actual track.
	
	According to the study by \citet{bi:6}, for the Whistler track, when the track distance is approximately 100m, the centrifugal force of the actual track exceeds the maximum allowable value. Although the track slope does not exceed the design limit of 0.204, a higher average slope, shorter total length, and larger height difference contribute to the centrifugal force surpassing the permissible threshold of 5g. In the case of the Cesana track, although its maximum design slope is 0.183, the actual slope exceeds 0.2, which may cause the centrifugal force during turns to exceed the allowable limit. The maximum centrifugal force of the actual track is about 4.5g, which remains within a reasonable range owing to the appropriate height difference and sufficient total track length. Although a smaller slope is critical for safety, an excessively small slope may reduce performance differences among athletes, thereby narrowing performance gaps.
	
\begin{figure}
	\centering
	\subfloat[Whistler track.]{%
		\resizebox*{10cm}{!}{\includegraphics{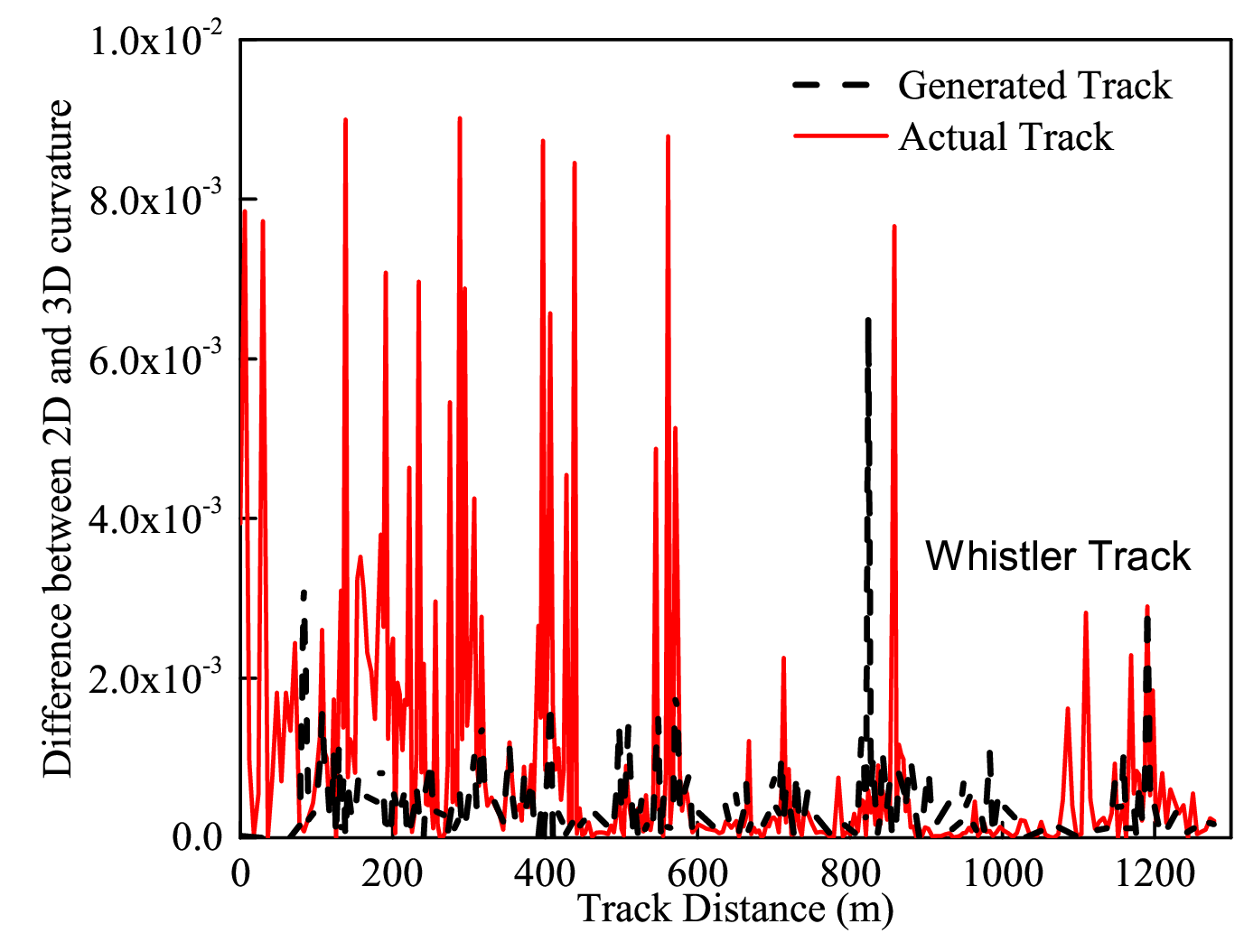}}}\hspace{5pt}
	\subfloat[Cesana track.]{%
		\resizebox*{10cm}{!}{\includegraphics{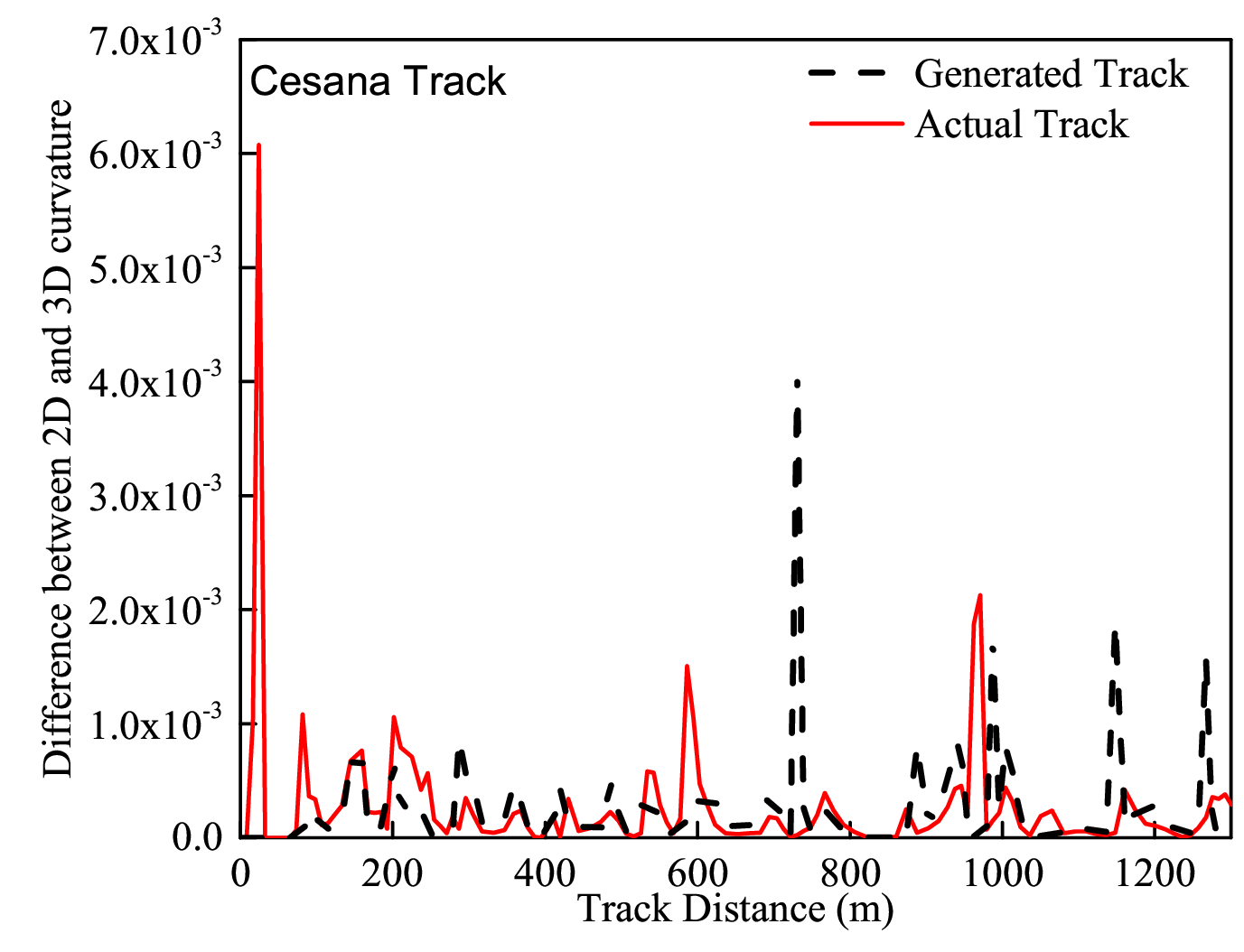}}}
	\caption{The comparison of trajectory differences between the generated and actual track centerlines.} \label{figure4}
\end{figure}

	Figures~\ref{figure4} shows the comparison of trajectory differences between the generated and actual track centerlines, in terms of 2D and 3D curvature variation along the track distance.  Among them, Figure~\ref{figure4}(a) is the Whistler track, and Figure~\ref{figure4}(b) is the Cesana track. As shown in the figures, in most regions, the differences between the 2D and 3D curvature of the generated track are smaller than or close to those of the actual track. This prevents excessive fluctuations in the slope in the z-direction and mitigates large vertical accelerations, thereby improving the safety and smoothness of the track design.
	
	 In summary, the proposed optimization equation can generate track centerlines based on real 2D data, and the proposed algorithm can compute centerlines whose characteristic parameter trends are consistent with those of the actual track. Simulation results on two selected tracks show that the length, height difference, and average gradient of the generated tracks are close to those of the actual tracks. Moreover, the slopes remain within the specified design limits, thereby validating the effectiveness of the proposed track centerline generation algorithm. The generation results are primarily influenced by parameters such as height difference, distance, average gradient, maximum gradient, average differences between the 2D and 3D curvature. In addition, the two real tracks analyzed herein highlight potential directions for further optimization.
	
	\subsection{Tracks Proportionally Scaled from Real 2D Planar Data}
	As real bobsleigh tracks with complete 2D planar data are scarce in the studies, while proportionally scaled versions of such data are relatively more available, proportionally scaled tracks are used here to validate the proposed track centerline generation model. By manually scaling real track data, this study simulates the case where original 2D data are unavailable, but proportionally scaled data are accessible. With the introduction of a scaling factor, the optimization equation is solved accordingly, and the generated track centerlines are compared with those of the actual tracks in terms of characteristic parameters.
	
	Taking the Whistler track as an example in this section, the optimization problem considers a scaling factor $f$, where the horizontal $x$ and $y$ coordinates of the Whistler track are uniformly scaled to 90\% of their original values (isotropic scaling is applied, with no inconsistency between the scaling ratios of $x$ and $y$ coordinates).
	
\begin{figure}
\centering
\resizebox*{13cm}{!}{\includegraphics{ 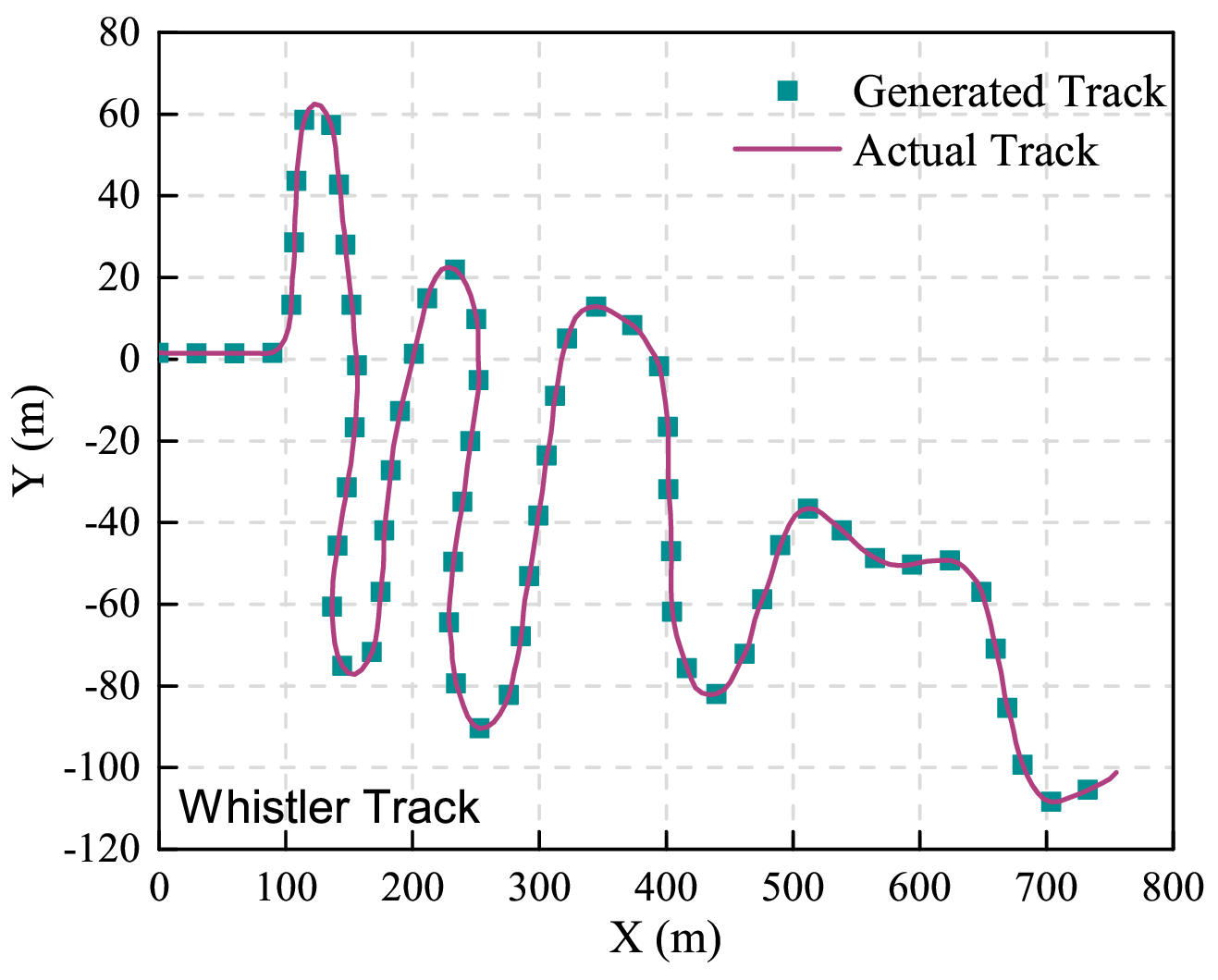}}\hspace{5pt}
\caption{ The comparison between the generated and actual $x$ and $y$ coordinates of the track centerlines.} \label{figure5}
\end{figure}

	Figure \ref{figure5} presents the comparison between the generated and actual $x$ and $y$ coordinates of the track centerlines. As shown in figure \ref{figure5}, the $x$ and $y$ coordinates calculated by the proposed algorithm follow the same trend as the actual values. The computed scaling factor $f$ is 1.1126, corresponding to a relative error of 0.14\% (with respect to $f$ = 1.1111). The result demonstrates that the proposed algorithm can effectively estimate the original $x$ and $y$ coordinates of the track centerline prior to scaling.
	
    To further analyze the computational results, Table \ref{table2} presents a comparison between the geometric parameters of the generated and actual track centerlines. As shown in Table \ref{table2}, Simulation results show that the length, height difference, and average gradient of the generated tracks are close to those of the actual tracks, with maximum deviations of 0.87\%, 3.8\%, and 3.6\%, respectively. Similar to Section 3.1, the generated tracks exhibit smaller average differences between the 2D and 3D curvature compared with their actual counterparts, while the slope values remain within the defined design limits.
	
		\begin{table}
		\caption{Comparison of geometric parameters of the generated and actual Whistler track.}
			\resizebox{\textwidth}{!}
		{\begin{tabular}{cccccc} 
				\toprule 
				& Total Length (m) & Height Difference (m) & Average difference between 2D and 3D curvature & Max Slope & Average Slope \\ 
				\midrule 
				Generated Track (Whistler) & 1289.1 & 154.9 & 4.17\texttimes10\textsuperscript{-4} & 0.1710 &  0.1218\\
				Actual Track (Whistler) & 1278 & 149.3 & 9.76\texttimes10\textsuperscript{-4} &0.204  & 0.1176  \\ 
				\bottomrule
		\end{tabular}}
		\label{table2}
	\end{table}

    In summary, the results obtained after applying the scaling factor further demonstrate the robustness and generalizability of the proposed optimization equation. Although minor deviations arise due to the scaling process, the overall geometric characteristics of the generated centerlines remain consistent with those of the actual tracks. The trends of the characteristic parameters—such as height difference, total length, and average gradient—are preserved, and the slopes stay within the defined design limits. These findings indicate that the proposed track centerline generation algorithm maintains accuracy and stability under scaled conditions, thereby confirming its applicability to tracks of varying dimensions. Moreover, the analysis of the scaled cases provides valuable insights for improving the parameter selection and enhancing the adaptability of the algorithm.

\section{Influence of Different Parameters on Track Generation}\label{sec:4}

In this section, tracks with real 2D centerline data (Whistler and Cesana tracks) and scaled 2D data (Nagano and Pyeongchang tracks) are selected as case studies to analyze the effects of segment count and weighting factors on track generation under the condition of actual and scaled 2D plane data. And the number of track segments and learning rate are determined separately for each track.

\subsection{Influence of the Number of Track Segments}

\begin{figure}
\centering
\resizebox*{15cm}{!}{\includegraphics{ 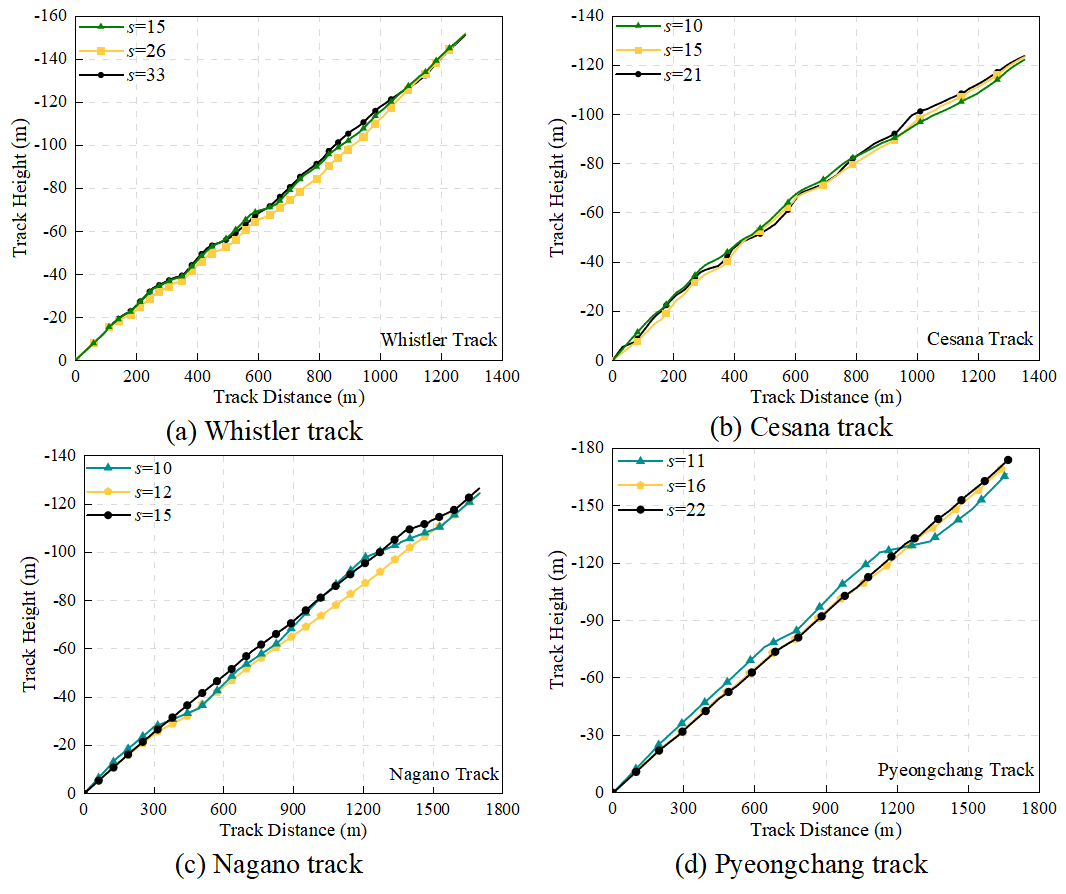}}\hspace{5pt}
\caption{ Influence of the number of track segments on track height variation along the track distance.} \label{figure6}
\end{figure}

	Figure~\ref{figure6} is influence of the number of track segments on track height variation along the track distance. As shown in figure~\ref{figure6}, the proposed method reproduces the global geometric characteristics of each track with small deviations, demonstrating stable optimization performance under different segmentation conditions. Simulation results for all four tracks indicate that the maximum deviation ranges for length, height difference, and average gradient are 0.1\%-1.1\%, 1.3\%-13.6\% and 1.1\%-10.7\%, respectively, with detailed values provided in Appendix A (Table \ref{tableA1}–\ref{tableA4}). All slope values remain within the specified design limits, confirming the robustness of the optimization framework under different weighting conditions.
	
	Increasing the number of segments primarily provides greater local flexibility, allowing more detailed adjustments of the segment-wise gradients. However, the effect of segmentation varies among tracks. For tracks with real 2D centerline data, show greater sensitivity to segmentation, as finer divisions better capture their local slope changes. In contrast, for tracks using scaled 2D data are less affected by segmentation count, showing stable results across different settings.
	
	Overall, these results suggest that the sensitivity to segmentation differs across tracks, and that selecting an appropriate number of track segments can effectively reduce deviations between the generated and actual track data, improving the geometric accuracy of the reconstructed centerlines.

\subsection{Influence of the Weighting Factor}

	The weighting factor of the height-difference term determines the degree to which the optimization prioritizes vertical conformity between the generated and actual tracks. In this section, its influence is analyzed across four representative tracks under varying weighting values (0.5, 1.0, and 1.5). Detailed values are provided in Appendix A (Table \ref{tableA5}–\ref{tableA8})
	
	Figure~\ref{figure7} is influence of the weighting factor on track height variation along the track distance. As shown in figure~\ref{figure7}, Across all cases, the proposed algorithm produces track centerlines whose length, height difference, and average gradient closely match those of the actual tracks, with maximum deviation ranges of 0.2–0.7\%, 1.0–16.4\%, and 0.8–13.5\%, respectively. Similar to the results obtained for the number of track segments, all slope values remain within the specified design limits.

	However, the sensitivity to the height-difference weighting varies among tracks. For tracks with real 2D centerline data, increasing the weighting factor enhances the alignment of the generated elevation profile with the actual track, as the optimization places greater emphasis on matching the total height difference. In contrast, for tracks using scaled 2D data, the effect of the weighting factor is less pronounced, since their geometric variation is already constrained by the scaling ratio. Consequently, the influence of the height-difference weighting is track-dependent, and selecting an appropriate weighting factor can effectively reduce deviations between the generated and actual track data, further improving the overall accuracy of the reconstructed 3D centerlines.

\begin{figure}
	\centering
	\resizebox*{15cm}{!}{\includegraphics{ 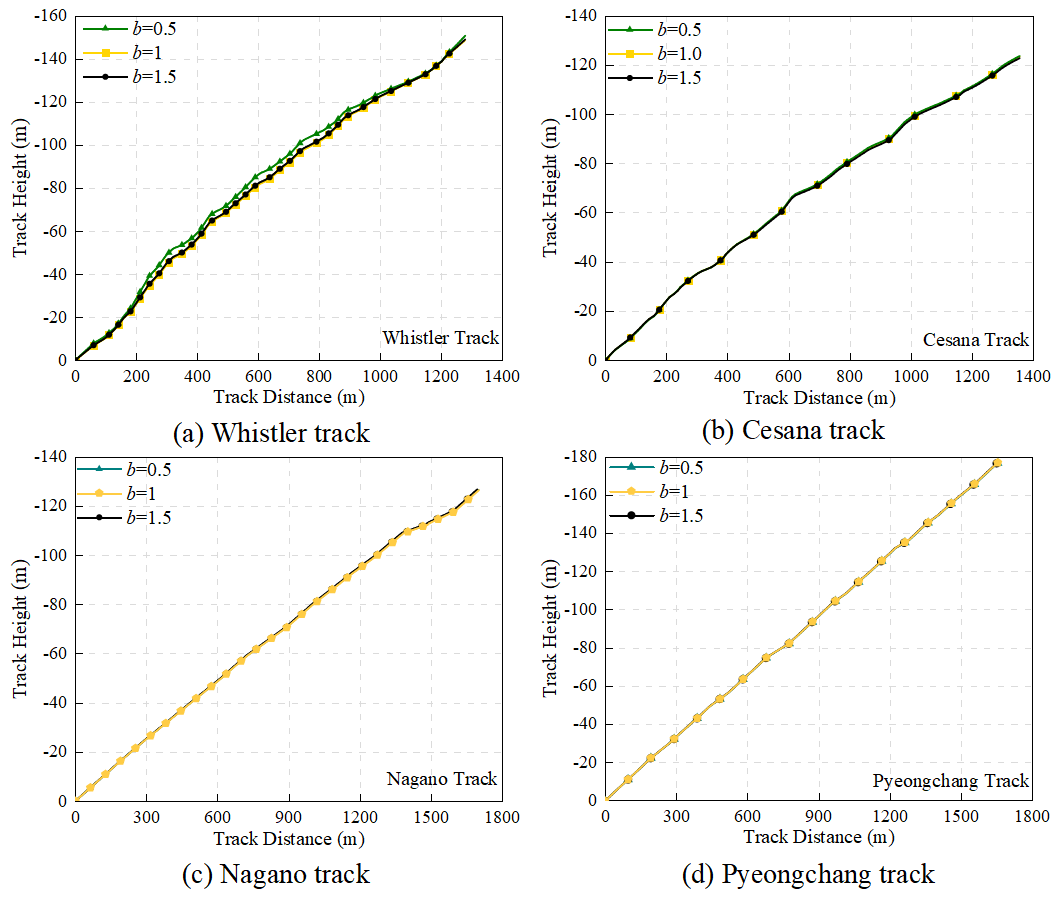}}\hspace{5pt}
	\caption{ Influence of the weighting factor on track height variation along the track distance.} \label{figure7}
\end{figure}

\section{Conclusions and future work}

This study proposes a method for generating a 3D bobsleigh track centerline based on a 2D centerline. In accordance with international track design regulations, the method considers factors such as total track length, height, and slope continuity to establish a track generation optimization problem.
﻿
Within the selected range of tracks in this study, the proposed algorithm can compute centerlines whose characteristic parameter trends are consistent with those of the actual track, based on either actual or scaled 2D data. Compared with the actual track, the generated centerlines are close to the actual values in terms of length, height difference, and average slope. Simulation results for all four tracks indicate that the maximum deviation ranges for length, height difference, and average gradient are 0.1\%-1.1\%, 1.0\%-16.4\%, and 0.8\%-13.5\%, respectively.

The sensitivity to segmentation and height-difference weighting differs across tracks, for tracks with real 2D centerline data, higher sensitivity to segmentation and weighting factors is observed, as finer divisions better capture local slope variations, and increasing the weighting factor improves the alignment between the generated and actual elevation profiles. In contrast, tracks based on scaled 2D data exhibit lower sensitivity to both segmentation count and weighting factor, maintaining consistent results under different configurations. Selecting an appropriate number of track segments and weighting factors can effectively reduce deviations between the generated and actual track data, improving the geometric accuracy of the reconstructed centerlines.

In the future, based on the wide range of generated national track data, the algorithm proposed in this study can be used to generate stylistically diverse bobsleigh tracks including track sections. Further work can focus on analyzing and constructing representative simulation tracks that incorporate key features from official tracks around the world. This will allow insights obtained from daily training, such as athletes’ driving strategies, to be more universally applicable across different official tracks worldwide. Moreover, exploring and comparing alternative optimization approaches including heuristic, stochastic, and learning-based methods may provide deeper insights into the trade-offs between solution quality, convergence speed, and robustness, relative to the current algorithm based framework.

\section{Declaration of conflicting interests}

The authors declared no potential conflicts of interest with respect to the research, authorship, and/or publication of this article.

\section{Funding}

The authors disclosed receipt of the following financial support for the research, authorship, and/or publication of this article: This work is supported by the Research on Open-Loop Dynamic Analysis Methods for Full-Vehicle Bobsleigh Systems of FAW R\&D (Grant No. KH54633401).

\section*{Data availability}

The data that support the findings of this study are available from the corresponding author, Jicheng Chen, upon reasonable request.

\bibliographystyle{apacite}
\bibliography{interactapasample}

\appendix
\section{Calculation results}

		\begin{table}[htbp]
			\caption{Comparison of geometric parameters of the generated and actual Whistler track under different numbers of track segments.}
				\resizebox{\textwidth}{!}{
				\begin{tabular}{cccccc} 
					\toprule 
					& Total Length (m) & Height Difference (m) & Average difference between 2D and 3D curvature & Max Slope & Average Slope \\ 
					\midrule 
					s=15 & 1286.8 & 147.3 & $4.52\times10^{-4}$ & 0.1703 & 0.1198\\
					s=26 & 1286.7 & 148.2 & $3.79\times10^{-4}$ & 0.156 & 0.1187\\ 
					s=33 & 1286.7 & 148.7 & $3.97\times10^{-4}$ & 0.152 & 0.1182\\
					Actual Track (Whistler) & 1278 & 149.3 & $9.76\times10^{-4}$ & 0.204 & 0.1176  \\ 
					\bottomrule
			\end{tabular}}
			\label{tableA1}
		\end{table}
		
				\begin{table}[htbp]
			\caption{Comparison of geometric parameters of the generated and actual Cesana track under different numbers of track segments.}
				\resizebox{\textwidth}{!}
			{\begin{tabular}{cccccc} 
					\toprule 
					& Total Length (m) & Height Difference (m) & Average difference between 2D and 3D curvature & Max Slope & Average Slope \\ 
					\midrule 
					s=10 & 1356.5 & 122.2 & 1.94\texttimes10\textsuperscript{-4} & 0.1425 &  0.0939\\
					s=15 & 1356.4 & 123.3 & 1.91\texttimes10\textsuperscript{-4} & 0.125 & 0.0926  \\ 
					s=21 & 1356.7 & 123.7 & 2.52\texttimes10\textsuperscript{-4} & 0.147 &  0.0920\\
					Actual Track (Cesana) & 1359.2 & 124.5 & 5.04\texttimes10\textsuperscript{-4} &0.183  & 0.0910\\ 
					\bottomrule
			\end{tabular}}
			\label{tableA2}
		\end{table}
		
		\begin{table}[htbp]
			\caption{Comparison of geometric parameters of the generated and actual Nagano track under different numbers of track segments.}
				\resizebox{\textwidth}{!}
			{\begin{tabular}{cccccc} 
					\toprule 
					& Total Length (m) & Height Difference (m) & Average difference between 2D and 3D curvature & Max Slope & Average Slope \\ 
					\midrule 
					s=10 & 1702.0 & 124.5 & 2.70\texttimes10\textsuperscript{-4} & 0.1057 &  0.0734\\
					s=12 & 1701.5 & 124.6 & 1.95\texttimes10\textsuperscript{-4} & 0.0867 & 0.0734 \\ 
					s=15 & 1699.7 & 126.5 & 2.04\texttimes10\textsuperscript{-4} & 0.0878 &  0.0746\\
					Actual Track (Nagano) & 1700 & 118 & - &0.150  & 0.0700\\ 
					\bottomrule
			\end{tabular}}
			\label{tableA3}
		\end{table}

		\begin{table}[htbp]
			\caption{Comparison of geometric parameters of the generated and actual Pyeongchang track under different numbers of track segments.}
			\resizebox{\textwidth}{!}
			{\begin{tabular}{cccccc} 
					\toprule 
					& Total Length (m) & Height Difference (m) & Average difference between 2D and 3D curvature & Max Slope & Average Slope \\ 
					\midrule 
					s=11 & 1662.0 & 166.5 & 2.47\texttimes10\textsuperscript{-4} & 0.1315 &  0.1008\\
					s=16 & 1648.6 & 170.3 & 1.87\texttimes10\textsuperscript{-4} & 0.1170 & 0.1038 \\ 
					s=22 & 1677.3 & 175.0 & 1.71\texttimes10\textsuperscript{-4} & 0.1150 &  0.1049\\
					Actual Track (Pyeongchang) & 1659 & 154 & - &0.25  & 0.0948\\ 
					\bottomrule
			\end{tabular}}
			\label{tableA4}
		\end{table}
		
	\begin{table}[htbp]
	\caption{Comparison of geometric parameters of the generated and actual Whistler track under different height-difference weighting factors.}
		\resizebox{\textwidth}{!}
	{\begin{tabular}{cccccc} 
			\toprule 
			& Total Length (m) & Height Difference (m) & Average difference between 2D and 3D curvature & Max Slope & Average Slope \\ 
			\midrule 
			b=0.5 & 1287.0 & 150.8 & 4.02\texttimes10\textsuperscript{-4} & 0.1434 &  0.1167\\
			b=1 & 1286.7 & 148.7 & 3.97\texttimes10\textsuperscript{-4} & 0.152 &  0.1182\\
			b=1.5 & 1286.7 & 149.1 & 3.96\texttimes10\textsuperscript{-4} & 0.157 & 0.1180  \\ 
			Actual Track (Whistler) & 1278 & 149.3 & 9.76\texttimes10\textsuperscript{-4} &0.204  & 0.1176  \\ 
			\bottomrule
	\end{tabular}}
	\label{tableA5}
\end{table}

\begin{table}[htbp]
	\caption{Comparison of geometric parameters of the generated and actual Cesana track under different height-difference weighting factors.}
		\resizebox{\textwidth}{!}
	{\begin{tabular}{cccccc} 
			\toprule 
			& Total Length (m) & Height Difference (m) & Average difference between 2D and 3D curvature & Max Slope & Average Slope \\ 
			\midrule 
			b=0.5 & 1356.3 & 122.8 & 1.60\texttimes10\textsuperscript{-4} & 0.1237 &  0.0922\\
			b=1 & 1356.7 & 123.4 & 1.62\texttimes10\textsuperscript{-4} & 0.1252 &  0.0926\\
			b=1.5 & 1356.4 & 123.7 & 1.63\texttimes10\textsuperscript{-4} & 0.1265 & 0.0927  \\ 
			Actual Track (Cesana) & 1359.2 & 124.5 & 5.04\texttimes10\textsuperscript{-4} &0.183  & 0.0910\\ 
			\bottomrule
	\end{tabular}}
	\label{tableA6}
\end{table}

\begin{table}[htbp]
	\caption{Comparison of geometric parameters of the generated and actual Nagano track under different height-difference weighting factors.}
		\resizebox{\textwidth}{!}
	{\begin{tabular}{cccccc} 
			\toprule 
			& Total Length (m) & Height Difference (m) & Average difference between 2D and 3D curvature & Max Slope & Average Slope \\ 
			\midrule 
			b=0.5 & 1702.6 & 126.4 & 2.02\texttimes10\textsuperscript{-4} & 0.0875 &  0.0744\\
			b=1 & 1699.7 & 126.5 & 2.04\texttimes10\textsuperscript{-4} & 0.0878 &  0.0746\\
			b=1.5 & 1695.4 & 126.7 & 2.06\texttimes10\textsuperscript{-4} & 0.0883 & 0.0749 \\ 
			Actual Track (Nagano) & 1700 & 118 & - &0.150  & 0.0700\\ 
			\bottomrule
	\end{tabular}}
	\label{tableA7}
\end{table}

\begin{table}[htbp]
	\caption{Comparison of geometric parameters of the generated and actual Pyeongchang track under different height-difference weighting factors.}
		\resizebox{\textwidth}{!}
	{\begin{tabular}{cccccc} 
			\toprule 
			& Total Length (m) & Height Difference (m) & Average difference between 2D and 3D curvature & Max Slope & Average Slope \\ 
			\midrule 
			b=0.5 & 1664.5 & 178.7 & 1.81\texttimes10\textsuperscript{-4} & 0.1172 &  0.1075\\
			b=1 & 1664.1 & 177.8 & 1.81\texttimes10\textsuperscript{-4} & 0.1172 &  0.1074\\
			b=1.5 & 1659.7 & 177.6 & 1.82\texttimes10\textsuperscript{-4} & 0.1174 & 0.1076 \\ 
			Actual Track (Pyeongchang) & 1659 & 154 & - &0.25  & 0.0948\\ 
			\bottomrule
	\end{tabular}}
	\label{tableA8}
\end{table}

\end{document}